\documentclass[12pt]{article}
\usepackage{geometry} 
\usepackage{moreverb,url}
\usepackage{hhline}
\usepackage{multirow}
\usepackage[colorlinks,bookmarksopen,bookmarksnumbered,citecolor=red,urlcolor=red]{hyperref}
\usepackage{xcolor}
\usepackage{amsfonts}
\usepackage{bbm}
\usepackage{soul}
\usepackage{amsmath}
\usepackage{booktabs}
\usepackage{mathtools}

\newcommand\BibTeX{{\rmfamily B\kern-.05em \textsc{i\kern-.025em b}\kern-.08em T\kern-.1667em\lower.7ex\hbox{E}\kern-.125emX}}

\newcommand\blfootnote[1]{%
  \begingroup
  \renewcommand\thefootnote{}\footnote{#1}%
  \addtocounter{footnote}{-1}%
  \endgroup
}

\date{\vspace{-2cm}}
\title{Beyond the classical type I error: Bayesian metrics for Bayesian designs using informative priors }

\begin{document}

\maketitle

\begin{center}
\author{Nicky Best\footnote[1]{GSK, UK}, Maxine Ajimi\footnote[2]{AstraZeneca, UK}, 
Beat Neuenschwander\footnote[3]{Novartis Pharma AG, Basel, Switzerland}, Gaëlle Saint-Hilary\footnote[4]{Saryga, France}\textsuperscript{,}\footnote[5]{Politecnico di Torino, Italy},\\Simon Wandel\footnote[3]{Novartis Pharma AG, Basel, Switzerland} on behalf of the PSI/EFSPI special interest group ``Historical Data"}
\end{center}

\blfootnote{Corresponding author: Nicky Best, GSK, 980 Great West Road, Brentford, Middlesex, TW8 9GS, UK. nicky.x.best@gsk.com}

\begin{abstract}
\noindent There is growing interest in Bayesian clinical trial designs with informative prior distributions, e.g. for extrapolation of adult data to pediatrics, or use of external controls. While the classical type I error is commonly used to evaluate such designs, it cannot be strictly controlled and it is acknowledged that other metrics may be more appropriate. We focus on two common situations -- borrowing control data or information on the treatment contrast -- and discuss several fully probabilistic metrics to evaluate the risk of false positive conclusions. Each metric requires specification of a design prior, which can differ from the analysis prior and permits understanding of the behaviour of a Bayesian design under scenarios where the analysis prior differs from the true data generation process. The metrics include the average type I error and the pre-posterior probability of a false positive result. We show that, when borrowing control data, the average type I error is asymptotically (in certain cases strictly) controlled when the analysis and design prior coincide. We illustrate use of these Bayesian metrics with real applications, and discuss how they could facilitate discussions between sponsors, regulators and other stakeholders about the appropriateness of Bayesian borrowing designs for pivotal studies.
\end{abstract}


\normalsize 

\section{Introduction}
The litmus test for new drugs are large, randomized clinical studies, known as pivotal trials. These trials are conducted to provide convincing evidence of a drug's efficacy and safety and to support its cost-effectiveness evaluation~\cite{Petrou2012}. The costs and the operational complexity of these studies are remarkable: in extreme cases, it can take hundreds of millions of dollars~\cite{Moore2018} and multiple years to conduct them. Unsurprisingly, this has led to the development of many statistical methods that aim to increase the efficiency, reduce the sample size or generally allow greater flexibility in the design of pivotal studies. Historically, most of these methods were formulated in the frequentist statistical framework. This is due to the dominating role that hypothesis testing (and the associated p-value) has been given in the assessment of novel therapies, with the common understanding that two adequately powered pivotal studies that show statistically significant results at a 2-sided $\alpha$-level of $0.05$ are required to establish effectiveness~\cite{FDAClinicalEvidence1998}. In this context, \textit{strict} type I error control has established itself as a guiding principle to judge whether the results of a trial can be used to contribute to the required level of evidence or not. Consequently, this precludes any method that formally incorporates existing evidence into the analysis of a pivotal trial, such as Bayesian designs with informative priors.

Over the last few years however, there has been increasing awareness of the substantial limitations which this strong focus on the frequentist framework and the type I error control entails. For example, in its recent draft guidance \textit{Demonstrating Substantial Evidence of Effectiveness for Human Drugs and Biological Products}~\cite{FDADemSubEvidence2019}, the US Food and Drug Administration (FDA) specifically discusses study designs that use external controls. Similarly, in its guidance \textit{Adaptive Design Clinical Trials for Drugs and Biologics}~\cite{FDAAdaptDesign2019}, the FDA highlights Bayesian designs that use informative priors. Furthermore, several approvals were granted both in the US and in Europe  based on non-randomized studies using external controls~\cite{Goring2019}. Even though these approvals were typically for rare diseases, they show the increasing willingness of regulators to consider evidence that has been generated outside the classical framework.

A particularly appealing approach to incorporate external evidence is via the use of informative prior distributions in the Bayesian framework. Since the prior distribution can be specified at the design stage, adherence to the principle of pre-specification is guaranteed, which ensures both transparency and avoidance of bias due to post-hoc decisions. Furthermore, through appropriate methods such as mixture priors~\cite{schmidli2014}, a possible prior-data conflict can also be mitigated at the design stage. Therefore, discussions and negotiations between the stakeholders (typically, regulators and pharmaceutical companies) can happen early in the process, i.e. before the study is started. Such discussions will concentrate on relevant aspects including, for example, the selection of the external data, the decision rule for claiming study success, and the metrics to evaluate the study design. Again, the FDA has published a draft guidance~\cite{FDAInteractingComplex2019} that summarizes in more granularity what information on the study design they consider important for such cases.

In descriptions of Bayesian designs with informative priors, however, it is common to find an evaluation of the associated classical (i.e.~frequentist or conditional) type I error. This type I error, if defined in the traditional way by considering only the sampling distribution of the observed data from the current trial, cannot be strictly controlled~\cite{viele2014, koppschneider2020, psioda2019} and depending on several factors, it can be above, below or equal to its nominal level. The question is of course whether this is a problem of the Bayesian approach, or of the metric itself. Intuitively, we might think that there must be a Bayesian error metric that bears the same essential property for Bayesian designs that the classical type I error does for frequentist designs, namely that it is controlled at a pre-specified level.

The goal of this paper is to investigate this question in more detail, while acknowledging that it matters whether information on one group (typically, the control), or on the treatment contrast, is borrowed. For these two situations, we will introduce several metrics, partially motivated by previous work~\cite{psioda2019, pennello2007}, and illustrate their use with real applications. Importantly, we prove that -- under certain conditions -- one particular metric, the average (i.e., unconditional) type I error, is actually controlled for Bayesian designs that leverage information on the control group. However, we believe that in their entirety, the metrics we propose fill a gap in the evaluation of Bayesian designs as emphasized e.g. in one of FDA's guidances~\cite{FDAInteractingComplex2019}.

The paper is structured as follows. First, we provide a short overview of the main ideas for constructing informative priors with a focus on the meta-analytic predictive prior. We then introduce the different metrics for Bayesian design evaluation. We illustrate them using two case studies, and conclude with a discussion.

\section{Methods}\label{sec:Methods}

\subsection{Introductory considerations}
We consider the case of a novel test treatment ($t$) being compared to a control treatment ($c$) in a pivotal study. We will refer to this pivotal study as the \textit{new} study. The underlying true treatment effects are denoted by $\theta_{t,\textit{new}}$ and $\theta_{c,\textit{new}}$, respectively. For example, $\theta_{c,\textit{new}}$ and $\theta_{t,\textit{new}}$ could be the (true) mean changes from baseline or the (true) log-odds of response under the test and control treatment. The key quantity of interest will be the treatment contrast, which, without loss of generality, we denote as $\delta_{\textit{new}} = \theta_{t,\textit{new}} - \theta_{c,\textit{new}}$. In the Bayesian framework, we quantify our information about the treatment contrast before the data from the new study are available by the prior distribution $p(\delta_{\textit{new}})$. Once the data $y_{\textit{new}}$ are available, the prior distribution will be updated according to Bayes theorem to obtain the posterior distribution, i.e.
\begin{eqnarray*}
p(\delta_{\textit{new}}|y_{\textit{new}}) \propto  p(y_{\textit{new}}|\delta_{\textit{new}}) \times p(\delta_{\textit{new}}) 
\end{eqnarray*}
We note that the above involves an explicit formulation for the prior of $\delta_{\textit{new}}$ in the sense that a  probability distribution for $\delta_{\textit{new}}$ is specified, and that the likelihood is also for the treatment contrast (such as a difference in means between treatment and control with corresponding standard error). However, this posterior distribution could also be derived in an implicit way by updating a prior distribution for $p(\theta_{c,\textit{new}}, \theta_{t,\textit{new}})$ to obtain a posterior $
p(\delta_{\textit{new}} = \theta_{t,\textit{new}} - \theta_{c,\textit{new}}|y_{t,\textit{new}}, y_{c,\textit{new}})$.

In practical applications, the choice of an explicit or implicit formulation will predominantly be taken based on whether information about the treatment effect(s) $\theta_{t,\textit{new}}, \theta_{c,\textit{new}}$, or the treatment contrast $\delta_{\textit{new}}$ is available.

We are now interested in leveraging information $y_h$ from one or more previous studies which will inform us about the (true) treatment effect of the control group or the (true) treatment contrast. While a number of approaches exist for leveraging this information, here we will focus on robust meta-analytic predictive priors as described in Schmidli et al  \cite{schmidli2014}. The robust meta-analytic predictive prior for parameter $\beta_{new} (= \theta_{c, new}$ or $\delta_{new}$ in the current setting) has the following form
\begin{eqnarray*}
p(\beta_{\textit{new}}|y_{h}) = \textit{w}\ p_{\textit{MAP}}(\beta_{\textit{new}}) + (1-\textit{w})\ p_{\textit{robust}}(\beta_{\textit{new}})
\end{eqnarray*}
where $p_{\textit{MAP}}(\beta_{\textit{new}})$ is the meta-analytic predictive prior estimated from the historical data, $p_{\textit{robust}}(\beta_{\textit{new}})$ the robust (vague, e.g., unit-information) prior, and $\textit{w}$ is the a priori weight. For further details regarding the derivation of the meta-analytic predictive prior, we refer to Schmidli et al \cite{schmidli2014}. As it is often overlooked, we note that the weights $\textit{w}, 1-\textit{w}$ will be updated to weights $\tilde{\textit{w}}, 1-\tilde{\textit{w}}$ once data are available, where the update follows mixture calculus \cite{schmidli2014}. 

Regardless of the situation, i.e., whether data on the control treatment or on the treatment contrast are leveraged, decision-making will typically be based on the posterior distribution of the treatment contrast. A typical success criterion 
takes the form
\begin{eqnarray}
\label{eq:success}
\textrm{Study success}  \equiv \mathbbm{1} \{ \textrm{Pr}(\delta_{\textit{new}} > \delta^{null}  |y_\textit{new}) \geq 1 - \alpha\}  
\end{eqnarray}
where $\mathbbm{1}$ is the indicator function. If $\delta^{null} = 0$, then this is a canonical success criterion for superiority, whereas for non-inferiority, $\delta^{null}$ will be a pre-specified boundary. 

While other success criteria have been discussed and are used in practice, too \cite{walley2015, roychoudhury2018}, in the following we restrict our considerations to the above definition. This success criterion leads to the same success region as a classical 1-sided test at level $\alpha$ in case that a truly non-informative (i.e., improper) prior for $\delta_{\textit{new}}$ is used.

\subsection{Metrics to evaluate Bayesian designs}
For simplicity, in the following we omit the subscript denoting historical or new study and use $\theta_t, \theta_c, \delta$, which will correspond to the parameters in the new study. Similarly, we also omit explicit conditioning on historical data in the prior distribution and just use $p(\theta_t), p(\theta_c)$ and $p(\delta)$, but when necessary we will make clear what information a prior depends on.

Classical frequentist type I error and power are well-established metrics to evaluate frequentist designs. Additionally, also assurance \cite{OHagan2005} is nowadays often used to contemplate design evaluations and acknowledge uncertainty. However, in fact all of these metrics are special cases of a common metric $m$, which, in the case of an explicit prior for the treatment contrast, can be expressed as
\begin{eqnarray}
\label{eq:general_metric_m}
m(\textrm{CP}(\delta), p(\delta)) =  \int \textrm{Pr}(\textrm{Study success} | \delta)p(\delta) d\delta
\end{eqnarray}
where 
\begin{eqnarray*}
\textrm{CP}(\delta) &=& \textrm{Pr}(\textrm{Study success} | \delta) \\
&=& \int_{y_\textit{new}} \mathbbm{1} \{ \textrm{Pr}(\delta_{\textit{new}} > \delta^{null} |y_\textit{new}) \geq 1 - \alpha\} p(y_\textit{new} | \delta) d y_\textit{new}
\end{eqnarray*}

 is simply the conditional power, which can be expressed as the expected value of (\ref{eq:success}) {\it wrt} the sampling distribution of the new study data given a fixed value for $\delta$ (and treating the historical data $y_h$ as fixed). 

A few interesting insights follow immediately. For example, the classical type I error is obtained by using a Dirac measure for $p(\delta)$ with point mass at $\delta^{null}$ (denoted $\Delta_{\delta^{null}}(.)$):
\begin{eqnarray}
m(\textrm{CP}(\delta), \Delta_{\delta^{null}}) &=&  \int \textrm{Pr}(\textrm{Study success} | \delta)\Delta_{\delta^{null}}(\delta) d\delta\nonumber \\
 &=& \underbrace{\textrm{Pr}(\textrm{Study success} | \delta=\delta^{null})}_{\textrm{Classical type I error}} \label{eq:typeIdelta}
\end{eqnarray}
Similarly, for the classical power we take a Dirac measure with point mass at the hypothesized value for the treatment contrast under the alternative. For assurance, we finally use a distribution for $p(\delta)$ that reflects our uncertainty around the hypothesized treatment effect. At this point, it becomes obvious that $p(\delta)$ is also a prior distribution. This has sometimes led to confusion in Bayesian designs, as there are now two priors present: the prior used for the actual analysis (which we will refer to as the \textit{analysis prior}), and the prior for the design evaluation (which we will refer to as the \textit{design prior}). This semantic differentiation should help to bring clarity with regards to which prior is the object of interest. Note that the analysis prior is not seen explicitly in \eqref{eq:general_metric_m}, nor in any of the subsequent equations in this section, as it is embedded in the definition of \textit{Study success}.

We note that from a very stringent viewpoint, the separation between \textit{analysis prior} and \textit{design prior} is not entirely adhering to the Bayesian paradigm: the \textit{analysis} prior is the best reflection of the evidence and the corresponding uncertainty and thus, this is the prior that should be used throughout. However, we acknowledge that calibration of Bayesian designs under different assumptions about the true parameter value(s) is useful \cite{grieve2016}, and is typically expected by regulatory agencies \cite{FDAInteractingComplex2019}. Indeed, \textit{drift} \cite{viele:2018,lim:2019} --- defined as the difference between the true parameter value in the new study and the historical prior value --- is the key quantity that drives bias (and hence the risk of incorrect conclusions) in Bayesian borrowing designs. Assumptions about how much drift is plausible therefore require careful consideration and discussion by the trial sponsor and regulatory agency. These assumptions could be reflected by fixing different values for $\delta$ as is done to evaluate classical type I error or power. Yet, which values to pick, and how likely are they? A natural extension is the aforementioned differentiation between the two priors, which facilitates the exploration of `what if' type of sensitivity scenarios for calibration purposes and has been advocated by several authors \cite{spiegelhalter2004,wang2002}. It essentially permits understanding of the behaviour of a Bayesian design under `drift' scenarios where the prior information used for analysis may differ from the true process generating the data.

When analysis prior information is available on $\theta_c$ and/or $\theta_t$, we typically wish to evaluate the design using design priors $p(\theta_c)$ and/or $p(\theta_t)$ instead of $p(\delta)$. In this case, it is convenient to also express the study success rule as a function of $\theta_t, \theta_c$ to make explicit the dependence on the design parameters of interest. This leads to the following formulation of metric (\ref{eq:general_metric_m}): 
\begin{eqnarray}
\label{eq:general_metric_control}
m(\textrm{CP}(\theta_c, \theta_t), p(\theta_c, \theta_t)) =  \int\int \textrm{Pr}(\textrm{Study success} | \theta_c, \theta_t) p(\theta_c, \theta_t) d\theta_c d\theta_t
\end{eqnarray}
Since the traditional definitions of type I error and power impose deterministic constraints on the relationship between $\theta_t$ and $\theta_c$ (i.e.~$\theta_t-\theta_c=\delta^*$ where $\delta^*$ equals the null or alternative value of the treatment contrast, respectively), it is immediately clear that a design prior for only {\it one} of $\theta_t$ {\it or} $\theta_c$ is required in order to evaluate metric (\ref{eq:general_metric_control}). If we assume a design prior $p(\theta_c)$, this leads to
\begin{eqnarray}
m(\textrm{CP}(\theta_c, \theta_t = \theta_c + \delta^*), p(\theta_c))  = \int \underbrace{\textrm{Pr}(\textrm{Study success} | \theta_c, \theta_t = \theta_c + \delta^*)}_{\textrm{Classical type I error or power}} p(\theta_c)d\theta_c 
\label{eq:typeIthetac}
\end{eqnarray}
Two interesting insights are as follows:
\begin{enumerate}
    \item  As already noted, for a Bayesian design with {\bf prior information only on the treatment contrast}, the metric defined by $m(\textrm{CP}(\delta), \Delta_{\delta^{null}})$ in \eqref{eq:typeIdelta} reduces to a {\bf single value} analogous to the classical type I error.
    \item For a Bayesian design with {\bf prior information on the control treatment}, the classical type I error (or power) is a {\it ``pointwise'' rate that depends on a single true value of $\theta_c$}, while the metric defined by $m(\textrm{CP}(\theta_c, \theta_t = \theta_c + \delta^\textrm{null}), p(\theta_c))$ in \eqref{eq:typeIthetac} is the {\bf average} (unconditional, or marginal) of this classical type I error with respect to the design prior distribution $p(\theta_c)$. 
 \end{enumerate}

\subsection{Specific considerations when borrowing information on the control response}
\label{sec:considerations_control}
For data generated under a normal likelihood, the average type I error defined in \eqref{eq:typeIthetac} is \textbf{strictly controlled} at level $\alpha$ if the \textit{analysis prior} is also used as the {\it design prior} $p(\theta_c)$ (see proof in supplementary material); in fact, for any distribution, it is at least asymptotically controlled at level $\alpha$. We find this an important and remarkable point, as it provides a bridge (and consistency) between two quite different worlds: in the frequentist world, we obtain strict type I error control as we are conditioning on the assumption of no treatment effect. In the Bayesian world, we obtain asymptotic (or, in some cases, strict) control as we are marginalizing over the \textit{analysis prior} distribution, assuming this is the best reflection of the available information. We think this is particularly re-assuring since it means once agreement has been reached on the \textit{analysis prior}, asymptotic type I error control with respect to the average (i.e., unconditional) type I error over the same prior (now viewed as the design prior) is guaranteed.

Instead of integrating the classical type I error or power over the design prior as in \eqref{eq:typeIthetac}, we may also consider a subset of the parameter space, or the maximum or minimum value over the range of the parameter space, for example:
\begin{eqnarray*} 
\underset{\theta_c \in \Theta_c}{\max}  \textrm{ Pr}(\textrm{Study success} | \theta_c, \theta_t = \theta_c + \delta^*)
\end{eqnarray*}
We also note that it may be of interest to evaluate the conditional power at true values other than the point null $\delta_{\textrm{null}}$. For example, we might be interested in the conditional power if the investigational treatment is assumed to be harmful, i.e. $\delta^* < \delta_{\textrm{null}}$. This concept will become important in the next section.

\subsection{Specific considerations when borrowing information on the treatment contrast} \label{sec:considerations_treat}

When borrowing on the treatment contrast, if the prior information supports a positive effect of the investigational treatment  (as is typically the case), this corresponds to a scenario where the prior is in conflict with the null treatment effect, resulting in an inflated classical type I error rate relative to what is typically considered acceptable. Kopp-Schneider et al \cite{koppschneider2020} and Psioda et al \cite{psioda2019} have shown that if strict control of type I error is required in this setting then prior information is effectively discarded. This is also reflected in FDA's guidance on Complex Innovative Trial Designs \cite{FDAInteractingComplex2019}, where it is stated that \textit{``when type I error probability is not applicable (e.g., some Bayesian designs that borrow external information), appropriate alternative trial characteristics should be considered''}.

We concur with Psioda et al \cite{psioda2020} that in cases where stakeholders believe reliable and relevant prior information on a treatment contrast is pertinent to the analysis of a new study, then to disregard this evidence in the name of ensuring strict type I error control is questionable. On the other hand, if relaxation of classical frequentist type I error control is to be permitted, other metrics are needed that help balance the potential efficiency gains of borrowing with unintended negative consequences that will arise when in reality, despite sound rationale, the prior information is not pertinent.  

One such alternative metric proposed by Psioda and Ibrahim \cite{psioda2019} is an average type I error that is somewhat analogous to \eqref{eq:typeIthetac}:
\begin{equation}
m(CP(\delta), p_{null}(\delta)) =\int\textrm{ Pr}(\textrm{Study success} | \delta) p_{null}(\delta) d\delta
\label{eq:typeInull}
\end{equation}
where the average is with respect to a design prior,  $p_{null}(\delta)$, which must be chosen to be consistent with the assumed null treatment effect. Psioda and Ibrahim \cite{psioda2019} suggest that a logical choice is to use the normalised analysis prior truncated on the range of values for the treatment effect that are consistent with the null:
\begin{equation} 
p_{null}(\delta) = \frac{p(\delta) \mathbbm{1} \{\delta \leq \delta^{null}\}}{\int_{\delta \leq \delta^{null}} p(\delta) d \delta} = \frac{p(\delta) \mathbbm{1} \{\delta \leq \delta^{null}\}}{\textrm{Pr}(\delta \leq \delta^{null})}
\label{eq:nullprior}
\end{equation}
%

Psioda and Ibrahim \cite{psioda2019} show that the average type 1 error \eqref{eq:typeInull} can be controlled at a specified level $\alpha$ through appropriate calibration of design parameters (e.g. prior weight on historical data and current trial sample size) implemented via a simulation-based grid search over the design space. They note, however, that in scenarios where the tail of the null design prior has negligible mass on values of $\delta$ inferior to $\delta^{null}$, then little or no historical information can be borrowed if the average type I error is to be controlled at conventional levels for $\alpha$. Importantly, for analysis priors specified directly on the treatment contrast, there is no equivalent of the result proved in the supplemental material for strong control of average type I error when borrowing on the control response. This is because, in situations where the analysis prior favours non-null treatment effects, there is a fundamental inconsistency between the analysis prior and the null treatment effect, and so the null design prior cannot be chosen to be consistent with the analysis prior. \\


Here, we propose an alternative metric which addresses the inconsistency between the prior information and the null treatment effect by explicitly accounting for the probability that the treatment effect is null or harmful under a suitably-chosen design prior:
\begin{footnotesize}
    \begin{eqnarray}
\tilde{m}(CP(\delta), \textrm{Pr} (\delta \leq \delta^{null}), p(\delta)) & = & 
\underbrace{m(CP(\delta), p_{null}(\delta))}_{\textrm{Average type I error}} \times \underbrace{\textrm{Pr}(\delta \leq \delta^{null})}_{\textrm{Prob treatment effect is null/harmful}} \nonumber \\
& = &\int\textrm{ Pr}(\textrm{Study success} | \delta)\frac{p(\delta) \mathbbm{1} \{\delta \leq \delta^{null}\}}{\textrm{Pr}(\delta \leq \delta^{null})} d\delta \times \textrm{Pr}(\delta \leq \delta^{null})  \nonumber \\
& = & \int\textrm{ Pr}(\textrm{Study success} | \delta) p(\delta) \mathbbm{1} \{\delta \leq \delta^{null}\} d\delta  \nonumber \\
& = & \int_{\delta \leq \delta^{null} }\textrm{ Pr}(\textrm{Study success} | \delta) p(\delta) d\delta
\label{eq:uncondFP}
\end{eqnarray}
\end{footnotesize}

%
%
%
Metric (\ref{eq:uncondFP}) is equal to the average type I error (\ref{eq:typeInull}) (where the average is \textit{wrt} the null design prior (\ref{eq:nullprior})) multiplied by the prior probability (under the corresponding untruncated version of the design prior $p(\delta)$) of the treatment effect being null or harmful. Closer inspection of the final row of equation (\ref{eq:uncondFP}) shows that this metric can also be interpreted as the joint probability of the true treatment effect being null or harmful {\it and} the study being declared a success. Spiegelhalter and Friedman \cite{spiegelhalter1986} call this the {\it type III error} of actually drawing a false positive conclusion, and it is sometimes also referred to as the {\it pre-posterior} probability of a false positive result. Metric (\ref{eq:uncondFP}) is also closely related to a metric proposed by Chuang-Stein and Kirby \cite{chuang-stein2017} to support decision-making for clinical trials. Their proposal is to calculate the probability of a correct decision for a trial design, which is the sum of two quantities: (1) the joint probability of the true treatment effect being beneficial {\it and} the trial being declared a success, plus (2) the joint probability of the true treatment effect being null or harmful {\it and} the trial failing to meet the success criteria. These various joint probabilities are illustrated in Table \ref{tab:jtprob}. It can be seen that the probability of a correct decision equals $p_{TP} + p_{TN}$, whilst our proposed metric in equation (\ref{eq:uncondFP}) to assess the pre-posterior probability of obtaining a false positive outcome is simply $p_{FP}$. 
\begin{table}[h]
\footnotesize\sf\centering
    \begin{center}
    \begin{tabular}{l|c|c}
        \toprule
Decision & \multicolumn{2}{c}{Truth} \\  \cline{2-3}
   & $\delta \leq \delta^{null}$ & $\delta >  \delta^{null}$ \\
   \hline
   	Study success  & $p_{FP}$  & $p_{TP}$ \\
    (accept $\delta > \delta^{null}$) & = Pr(false positive result) &  = Pr(true positive result)  \\
    \hline
   	Study failure  & $p_{TN}$  & $p_{FN}$   \\
    (accept $\delta \leq \delta^{null}$) & = Pr(true negative result) & = Pr(false negative result) \\
        \bottomrule
    \end{tabular}
    \caption{Joint probabilities of (truth, decision) for a clinical trial (modified from Chuang-Stein and Kirby \cite{chuang-stein2017}).} 
    \label{tab:jtprob}
    \end{center}
\end{table}

Calculation of (\ref{eq:uncondFP}) (and the other probabilities in Table \ref{tab:jtprob}) requires specification of a suitable design prior for $\delta$. As with the average type I error defined in (\ref{eq:typeIthetac}), the analysis prior may be used for this design prior, or other choices could be considered, such as a prior representing sceptical beliefs about the true treatment contrast. 

An upper bound on (\ref{eq:uncondFP}) can be obtained by replacing Pr(Study success $| \delta$) by Pr(Study success $| \delta^{null}$), where the latter is the classical type I error (\ref{eq:typeIdelta}). This reduces sensitivity of the metric (\ref{eq:uncondFP}) to how quickly the design prior down-weights values of $\delta$ that are more extreme than the point null, since only the maximum error rate (i.e. the traditional type I error when $\delta = \delta^{null}$) enters into the calculation:
\small
\begin{eqnarray}
\label{eq:upperbound}
\max\{\tilde{m}(\textrm{CP}(\delta), \textrm{Pr} (\delta \leq \delta^{null}), p(\delta))\} 
\!\!\! &=& \!\!\! \int_{\delta \leq \delta^{null} }\max_{\delta \leq \delta^{null}}\{\textrm{ Pr}(\textrm{Study success} | \delta)\} p(\delta) d\delta  \nonumber \\
\!\!\! &=& \!\!\! \textrm{ Pr}(\textrm{Study success} | \delta^{null}) \int_{\delta \leq \delta^{null} } p(\delta) d\delta \ \nonumber \\
\!\!\! &=& \!\!\! \underbrace{\textrm{ Pr}(\textrm{Study success} | \delta^{null})}_{\textrm{Classical type I error}} \times  \underbrace{\textrm{ Pr}(\delta \leq \delta^{null})}_{\mathclap{\substack{\text{Prob treatment effect} \\ \text{is null/harmful}}}} 
\end{eqnarray} 
\normalsize
Another way to derive metric (\ref{eq:upperbound}) is to evaluate metric (\ref{eq:uncondFP}) using a `spike and slab' design prior for the treatment contrast, i.e.~$p(\delta) = \pi_{null} \Delta_{\delta_{null}} + (1-\pi_{null}) p_{benefit}(\delta)$, where the prior probability density on null or harmful values is a Dirac measure concentrated on a point mass (the `spike') at $\delta^{null}$ with weight $\pi_{null}$ and $p_{benefit}(\delta)$ represents the design prior for non-null (beneficial) values of the treatment effect. Under this design prior, metric (\ref{eq:uncondFP}) becomes:

\small
\begin{equation*}
\label{eq:upperbound2}
\begin{split}
\tilde{m}(CP(\delta), \pi_{null}, \pi_{null} \Delta_{\delta_{null}} &+ (1-\pi_{null})p_{benefit}(\delta))   =  \\
& \underbrace{\int\textrm{ Pr}(\textrm{Study success} | \delta) \Delta_{\delta_{null}} d\delta}_{\textrm{Classical type I error}} 
\quad \times \quad \underbrace{\pi_{null}}_{\mathclap{\substack{\text{Prob treatment} \\ \text{effect is null}}}}  
\end{split}
\end{equation*}
\normalsize
which is the same as metric (\ref{eq:upperbound}).

\section{Case Study 1: Borrowing historical placebo data}
We illustrate several metrics on a case-study inspired by a double-blind, randomised, placebo-controlled proof-of-concept study testing whether secukinumab, a human anti-IL-17A monoclonal antibody, was safe and effective for the treatment of moderate to severe Crohn's disease \cite{Hueber2012, RBesT}. The disease status is assessed with the Crohn's Disease Activity Index (CDAI), which consists of a weighted sum of eight clinical or laboratory variables \cite{best1976}. CDAI scores can range from 0 to about 600, where an asymptomatic condition corresponds to a value below 150, and severe disease is defined as a value greater than 450. The study's primary endpoint is the CDAI change from baseline at week 6, with negative values indicating an improvement of the patient's condition. Historical placebo data are available from 6 studies with a total of 671 patients on placebo (see supplementary material, Table S1). The primary endpoint is assumed normally distributed with a known standard deviation of $\sigma=88$, estimated from the literature and historical studies.

\subsection{Design}
We discuss a Bayesian Dynamic Borrowing (BDB) design for the new study, using a 2:1 randomisation ratio (treatment to control) and an informative prior for the placebo arm to supplement the data from the concurrently randomised placebo subjects. The planned sample size is $40$ patients in the active treatment group and $20$ patients in the placebo group. Assuming a true difference of $-70$ in the change from baseline at 6 weeks as compared to placebo, the study provides $83\%$ power in a frequentist framework with 1-sided $\alpha = 0.025$. While the original study was designed with a dual-criterion \cite{roychoudhury2018}, we use here for simplicity a single Bayesian success criterion equivalent to classical statistical significance under an improper prior, which we define as:
$$\textrm{Pr}\left(\theta_{t, new}-\theta_{c, new} < 0\right) \geq 0.975,$$
where $\theta_{t, new}$ and $\theta_{c, new}$ denote the true mean CDAI change from baseline at week 6 in the active and control arms, respectively, in the new study.

\begin{center}
\begin{figure}[bt]
    \centering
    \includegraphics[scale = 0.75]{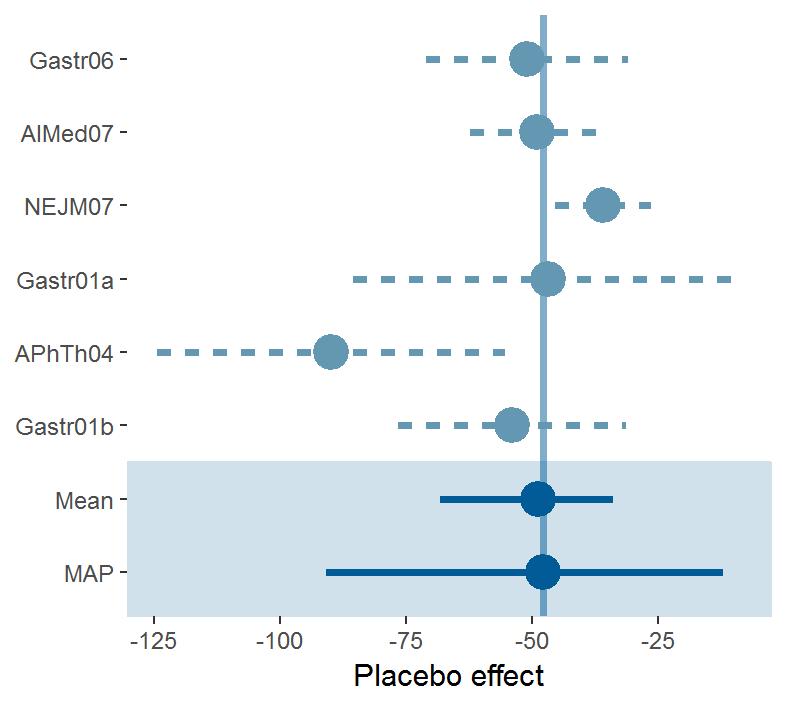} \includegraphics[scale = 0.75]{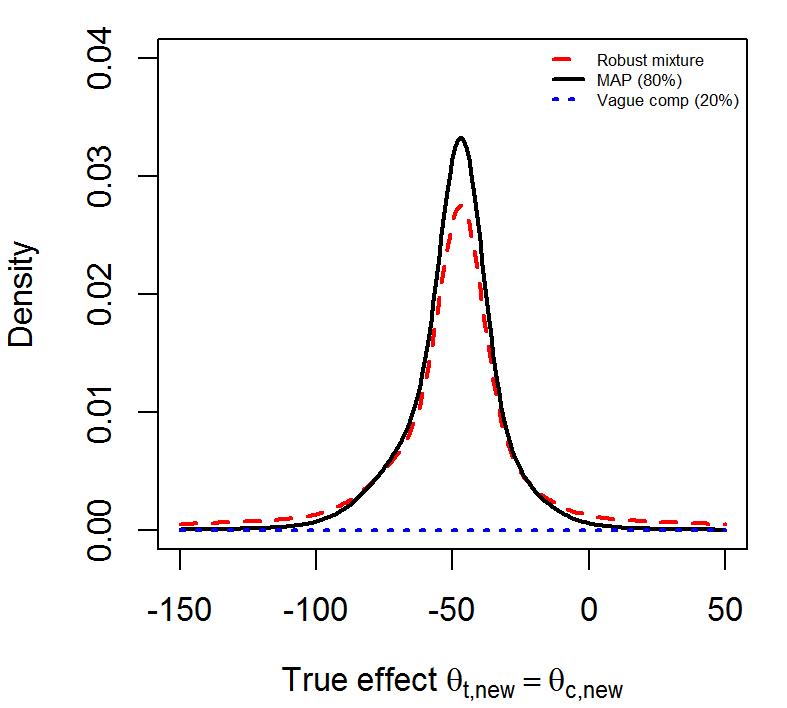}
    \caption{Crohn's disease application. Left: Forest plot of observed data for the historical studies, and posterior and predictive (MAP) estimates of the true placebo  effect $\mu_{c}$. Right: Robust MAP prior distribution with prior weights of 80\% on the historical data (MAP) and 20\% on the vague (robust) component.}
    \label{fig:Crohn_forest_rmap}
\end{figure}
\end{center}

We derive an informative prior for the control arm using the meta-analytic-predictive (MAP) approach \cite{schmidli2014, Neuenschwander2010} (see details in supplementary material), and approximate it with a mixture Normal distribution \cite{RBesT} with three components:
\begin{eqnarray}
p_{MAP}(\theta_{c,\textit{new}} \ \mid \ y_{c,1}, ..., y_{c,6}) = & 0.51 \times \mathcal{N}(-51.0, 19.9^2) \nonumber \\ 
& + 0.44 \times \mathcal{N}(-46.8, 7.6^2)\nonumber \\ 
& + 0.05 \times \mathcal{N}(-54.1, 51.7^2) \label{eq:MAPprior}
\end{eqnarray}
where $p_{MAP}$ denotes the density of the MAP prior. Finally, we robustify the MAP prior to hedge a possible prior-data conflict \cite{schmidli2014}, by adding a robust vague component to the mixture distribution (Figure~\ref{fig:Crohn_forest_rmap}):
\begin{align} 
p_{RMAP}(\theta_{c,\textit{new}} \ \mid \  y_{c,1}, ..., y_{c,6}) = &  \ 0.8 \times p_{MAP} \nonumber  \\ 
& \mathbf{+ \ 0.20 \times \mathcal{N}(-50.0, 88^2)}, \label{eq:rMAPprior} 
\end{align}
where $p_{RMAP}$ denotes the density of the robust MAP prior. The mean of the vague component is set to $-50$ to be consistent with the historical data, and its standard deviation is set to $\sigma=88$, so that the vague prior component is approximately equivalent to one subject’s worth of information. The choice of the weight of $20\%$ on the robust component reflects our prior belief about the possibility of non-exchangeability between the placebo effect estimated from historical studies and the placebo effect in the new study.

We consider three different analysis priors for the placebo arm:
\begin{itemize}
    \item A (very) vague prior, defined as $\theta_{c,\textit{new}}\ \sim \ \mathcal{N}(-50, 8800^2)$, not borrowing historical information
    \item The MAP prior (\ref{eq:MAPprior})
    \item The robust MAP prior (\ref{eq:rMAPprior})
\end{itemize}
The vague prior is also used for the active arm in all designs. 

\subsection{Classical type I error}
\begin{center}
\begin{figure}[bt]
    \centering
    \includegraphics[scale = 0.90]{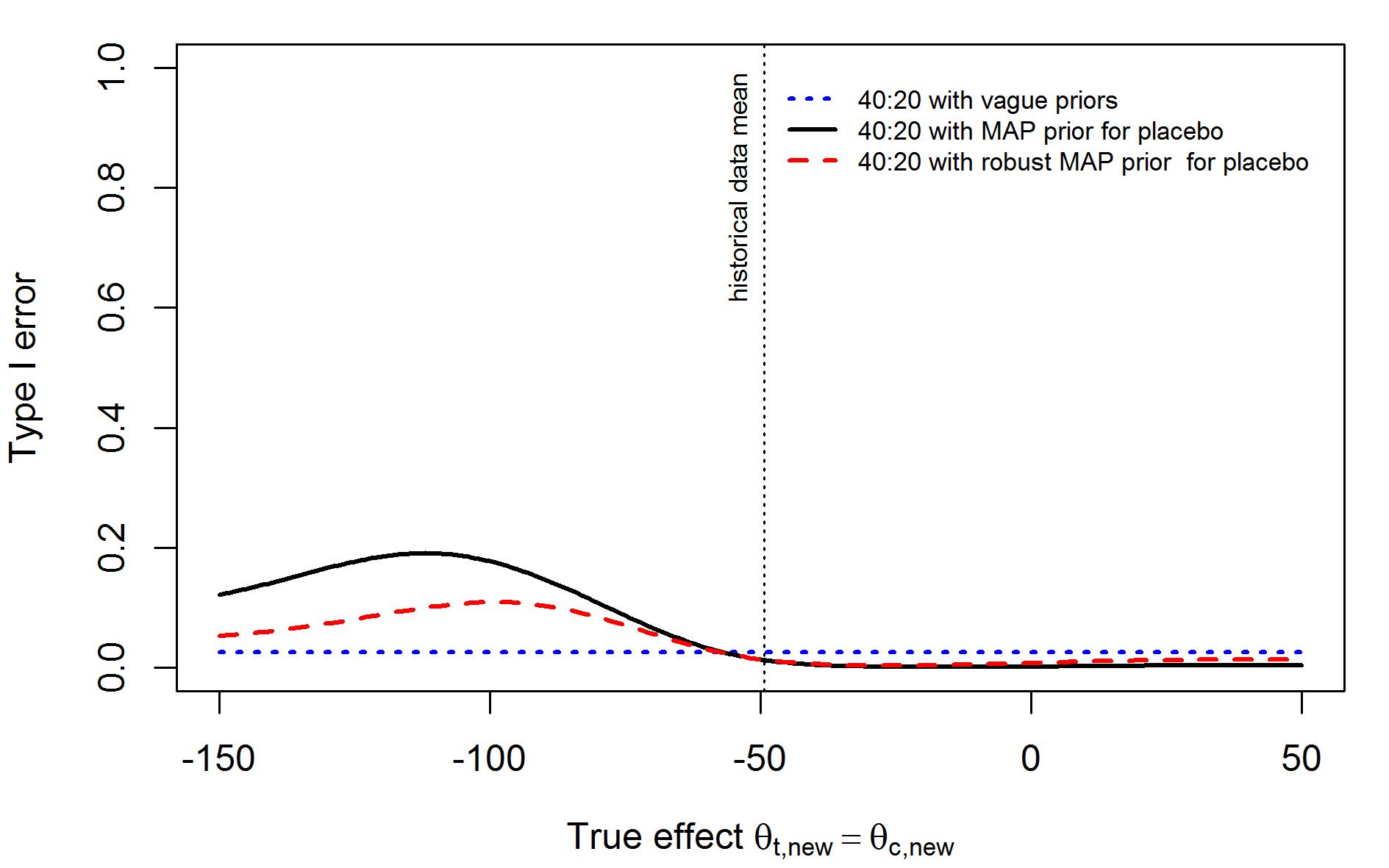} 
    \caption{Crohn's disease application: classical type I error for Bayesian designs with 3 different analysis priors for the control arm.}
    \label{fig:Crohn_type1}
\end{figure}
\end{center}

Figure~\ref{fig:Crohn_type1} presents the pointwise classical (frequentist) type I error rate for each of the three Bayesian designs over a large but plausible range of true values of CDAI mean change from baseline between $-150$ and $50$, when both the treatment and the placebo are assumed to have the same effect (i.e. $\theta_{t,new}$ = $\theta_{c, new}$). First, it can be seen that, when using vague priors for both arms (i.e.~no borrowing of historical data), the classical type I error rate does not depend on the true value of the placebo effect $\theta_{c, new}$ and is controlled at $2.5\%$. On the other hand, when borrowing historical placebo data (using either MAP or robust MAP prior), the pointwise classical type I error rate can be increased or decreased, depending on the difference between the true placebo effect in the new study and the observed effect in historical data. In particular, when the true placebo effect is better (more negative) than the observed effect in historical data, the classical type I error rate can be increased compared to the nominal level. In this case, using historical information penalises the observed placebo effect, leading to a more pronounced treatment contrast estimate and increasing the classical type I error. Over the range of values for the true placebo response shown in Figure~\ref{fig:Crohn_type1}, the maximum of the classical type I error rate is $19\%$ with the MAP prior, but it can theoretically reach $100\%$ for biologically implausible values of $\theta_{c, new}$ (lower than $-1000$, see supplementary material, Figure S1). As expected, the robust MAP prior reduces this type I error inflation to a maximum of $11\%$ on this range (and across the entire range), as the informative components in the prior are entirely discarded for large prior-data conflicts. Readers may also have noted that the type I error rates for both the MAP and robust MAP prior designs are slightly {\it lower} than the nominal level when the true placebo effect is identical or very similar to the observed historical value; this relates to the fact that the sampling distribution of the posterior test statistic (i.e.~the indicator of whether the posterior success criteria is met or not) is unbiased (or nearly unbiased) and has less variability (due to the prior information) than under an improper prior so the tail area will be less than $\alpha$.

\subsection{Average type I error}
\begin{center}
\begin{figure}[bt]
    \centering
    \includegraphics[scale = 0.80]{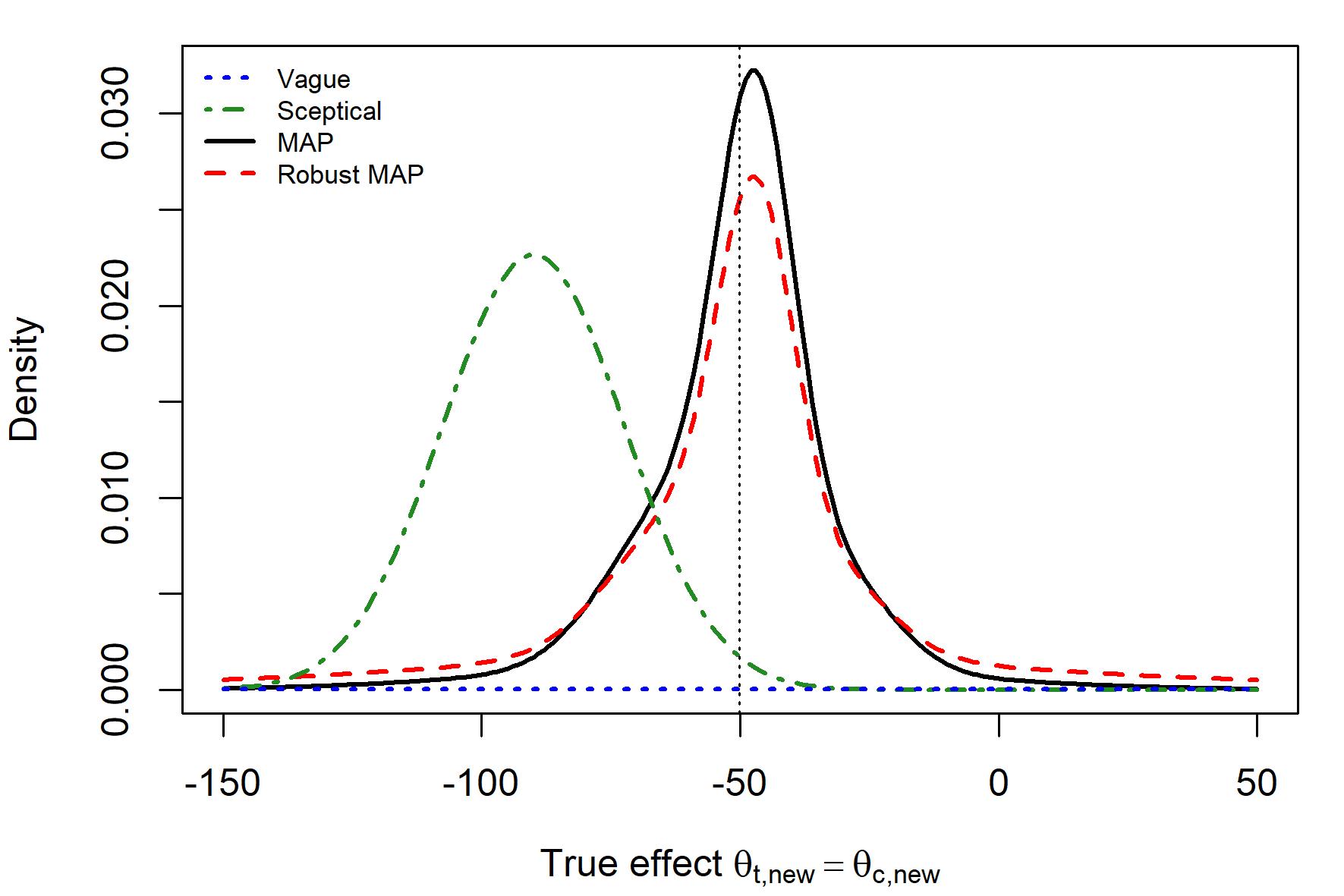} 
    \caption{Crohn's disease application. Different design priors used for the average type I error calculations.}
    \label{fig:Crohn_truedist}
\end{figure}
\end{center}

We evaluate the \textit{average} (unconditional) type I error proposed in (\ref{eq:typeIthetac}) over four design priors (see Figure~\ref{fig:Crohn_truedist}): 

\begin{enumerate}
    \item A ``vague'' design prior chosen to be the same as the vague prior used for the analysis.
    \item A ``sceptical'' design prior $\theta_{c,\textit{new}}\ \sim \ \mathcal{N}(-90, 25^2)$ that corresponds to the posterior distribution of a stand-alone analysis of the historical study with the ``most extreme'' placebo effect (APhTh04).
    \item A ``realistic" design prior based on all relevant historical data, chosen to be the same as the MAP analysis prior.
    \item A ``robust" design prior chosen to be the same as the robust MAP analysis prior.
\end{enumerate}
The results are presented in Table~\ref{tab:Crohn_results}. 
A consistent viewpoint is to use the same distribution as analysis prior and design prior, meaning that the distribution of assumed values for the placebo effect used to evaluate the design corresponds to the prior assumption about the placebo effect used in the analysis. Under this viewpoint the average type I error rate is controlled at $2.5\%$ (values in bold in Table~\ref{tab:Crohn_results}), as discussed in Section \ref{sec:considerations_control} and proved in supplementary material. As sensitivity analyses, we assume some scepticism about the distribution of the true placebo effect, to help understand how the analysis prior behaves when using another design prior. 

First, we observe that the average type I error is always controlled to its nominal level when the analysis prior is vague, regardless of which design prior is assumed. This is expected since there is no increase of the pointwise frequentist type I error rate with this analysis prior. 

On the other hand, when the BDB design is evaluated under a scenario where all possible values of the true placebo effect are considered (almost) equally likely (vague design prior), we observe increased average type I error rates ($48.5\%$ and $45.6\%$ for designs using the MAP or robust MAP analysis prior respectively). This is because extreme values for the true control effect are considered to be possible with the vague design prior, resulting in large, but implausible, prior data conflicts. 

The type I error rate is much lower when it is averaged over more informative design priors compared to the vague design prior. Assuming the sceptical design prior for the placebo effect results in an average type I error of $13.4\%$ with the MAP analysis prior, reduced to $8.8\%$ with the robust MAP analysis prior thanks to the robustification. Averaging over the plausible, but conservative, robust MAP design prior leads to a small type I error increase with the MAP analysis prior ($3.2\%$). On the other hand, the type I error is reduced to $2.2\%$ when using the robust MAP analysis prior but averaging over the MAP design prior. 

In summary, assuming the same distribution for the analysis and the design priors is the most consistent assumption, and our results provide empirical confirmation that the average type I error of BDB designs is controlled in these scenarios. Sensitivity analyses were conducted to assess the impact of evaluating the design using assumptions (design prior) for the true placebo effect that are not consistent with the analysis prior: they illustrate that average type I error increases are possible, but they are large only for highly implausible assumptions (vague design prior).   
\begin{table}[h]
\small\sf\centering
    \begin{center}
    \begin{tabular}{lcccc}
        \toprule
   \textbf{Analysis prior} & \multicolumn{4}{c}{\textbf{Design prior for placebo effect}} \\
   \textbf{for placebo effect} & Vague &	Sceptical	& 	MAP	& Robust MAP\\
   \hline
   	Vague	& \textbf{2.5\%}	& 2.5\%	& 2.5\%	& 2.5\% \\
	MAP	& \underline{48.5\%}	& \underline{13.4\%}	& \textbf{2.5\%}	& \underline{3.2\%} \\
	Robust MAP	& \underline{45.6\%}	& \underline{8.8\%}	& 2.2\%	& \textbf{2.5\%} \\
        \bottomrule
    \end{tabular}
    \caption{Crohn's disease application. Average type I error. Values for scenarios where the analysis prior matches the design prior are highlighted in bold, and values greater than $2.5\%$ are underlined.} 
    \label{tab:Crohn_results}
    \end{center}
\end{table}

\section{Case Study 2: Borrowing historical information on a treatment contrast}

The second case study is inspired by a recent FDA approval of sBLA 125370/S-064 for belimumab (Benlysta) IV formulation for use in children aged 5-17 years. Benlysta was approved by the FDA for adult patients with active, seropositive lupus erythematosus (SLE) in 2011. A paediatric post-marketing study was required and the applicant undertook to conduct a randomized, double-blind, placebo-controlled trial targeting to enroll 100 paediatric subjects 5 to 17 years of age with active systemic SLE. The paediatric study was not fully powered by design, efficacy was planned to be descriptive and no formal statistical hypothesis testing was proposed. The study was completed in 2018, with a total of 92 subjects. 

To facilitate the review of Benlysta, the FDA requested a post-hoc Bayesian analysis to further evaluate the efficacy of Benlysta in paediatric SLE patients by utilising relevant information from the adult studies. The rationale was to provide more reliable efficacy estimates in the pediatric study in a setting where the clinical review team believed that the disease and patient response to treatment are likely to be similar between adults and paediatrics (see FDA's review \cite{FDABLA} for details). For the purposes of the present (hypothetical) case-study, we consider how a study of Benlysta in paediatric subjects could have been prospectively designed using a {\it pre-specified} BDB analysis borrowing efficacy data from the adult pivotal trials to provide confirmatory evidence of a positive benefit-risk of Benlysta in children with SLE. 

\subsection{Design of paediatric trial}

Evidence of efficacy has been established in adults in two independent pivotal Phase 3 trials, which are pooled and considered to be one single source of historical data, indexed by $h$. The primary endpoint was response at week 52 on the SLE responder index (SRI) \cite{FDABLA}, and the summary measure of treatment effect was the odds ratio for Benlysta compared to placebo. The pooled odds ratio based on a total of $N_h = 1125$ subjects from these studies was $1.62$ (95\% CI $1.27$ – $2.05$), which on the log odds ratio scale corresponds to a point estimate of $y_h = 0.48$ with standard error of $s_h = 0.121$. 

A paediatric trial is proposed using a BDB design, to draw inference about the treatment effect in the paediatric population by supplementing the paediatric data with data from the pivotal adult studies. The planned sample size is 100 patients (50 patients per arm), with study success defined as having at least $97.5\%$ posterior probability that the true log odds ratio of response in paediatrics at week 52 on the SRI disease activity index on Benlysta compared to placebo exceeds $0$:
$$\textrm{Pr}\left(\delta_{new} > 0\right) \geq 0.975,$$
where $\delta_{new} $ denotes the true paediatric log odds ratio of response. 

\begin{center}
\begin{figure}[bt]
    \centering
    \includegraphics[scale = 0.60]{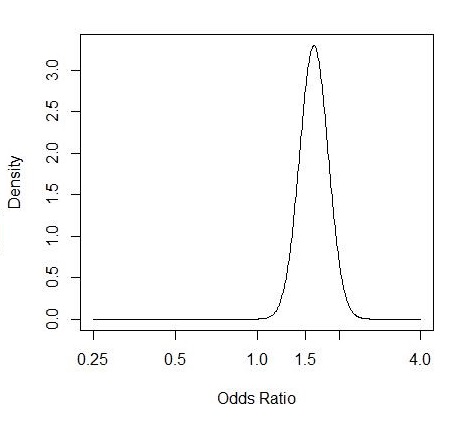} \includegraphics[scale = 0.6]{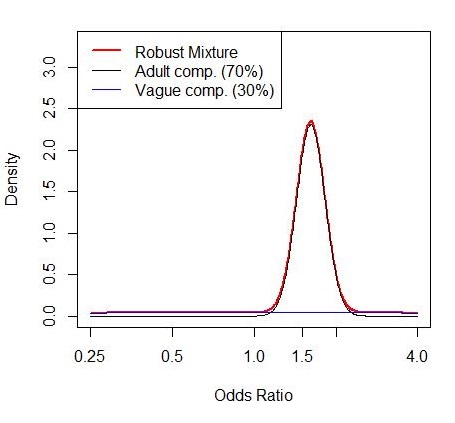}
    \caption{Paediatric lupus application. Left: Prior based on adult data (= posterior distribution of the odds ratio from the pooled adult studies, assuming an initial non-informative prior). Right: Robust mixture prior distribution with prior weights of 70\% on the adult data and 30\% on the vague (robust) component. }
    \label{fig:Benlysta_forest_rmap}
\end{figure}
\end{center}

Let denote $y_{new}$ the observed log odds ratio of response from a logistic regression of the paediatric data and $s_{new}$ its estimated standard error. We assume a Normal sampling distribution $ y_{new}  \sim \mathcal{N}(\delta_{new}, s^2_{new})$ with $s_{new}$ treated as fixed, and we build a Bayesian robust mixture prior for the paediatric treatment contrast $\delta_{new}$ (Figure \ref{fig:Benlysta_forest_rmap}):
\begin{eqnarray*}
 p(\delta_{new} \mid y_h, s_h) & = & 0.7 \times \mathcal{N}(0.48, 0.121^2) 
\\ && + \ 0.3 \times \mathcal{N}(0, 2.87^2). 
\end{eqnarray*}
The informative component represents the posterior distribution for the adult treatment contrast (log odds ratio) obtained from a Bayesian analysis of the pooled adult data $y_h$, assuming a Normal sampling distribution with known variance $s_h^2$ and an improper prior on the mean. A prior weight of $w = 70\%$ is assigned to this adult component, representing the prior degree of belief that the adult treatment effect estimated from the pivotal studies provides relevant information about the treatment effect in paediatric patients. The mean of the vague component is set to $0$ (i.e. centred at the null hypothesis of no effect) and the variance is set to $N_h/2 \times s^{2}_{h} = 2.87^2$ so that the effective sample size of the vague component is worth just one subject per arm.  

For comparison, we also consider a Bayesian analysis with a vague prior defined as $\delta_{new}  \sim \mathcal{N}(0, 100^2)$, not borrowing any adult information.

\subsection{Classical type I error}

\begin{center}
\begin{figure}[bt]
    \centering
    \includegraphics[scale = 0.31]{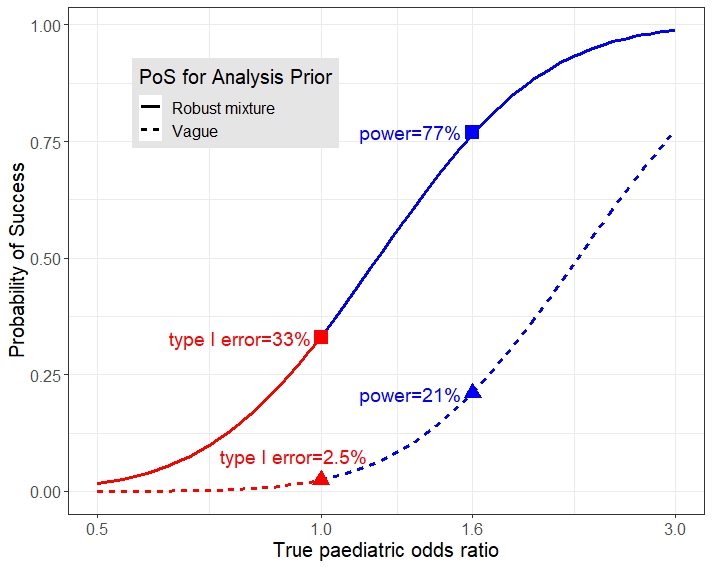} 
    \includegraphics[scale = 0.26]{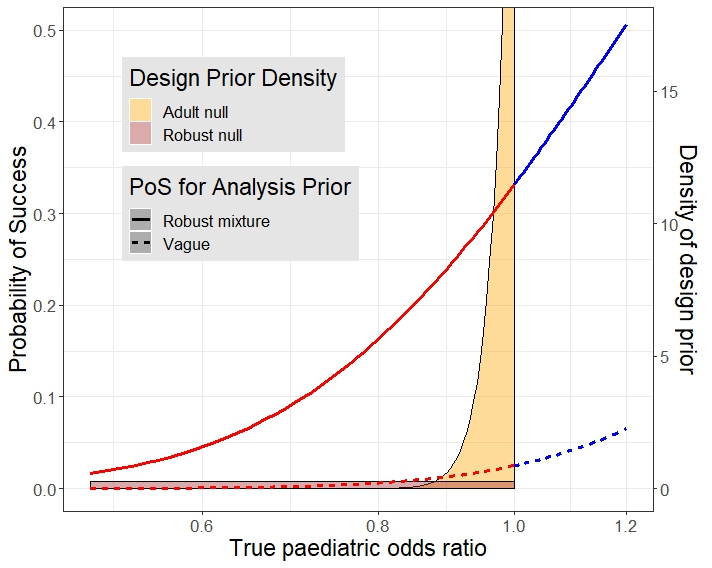} 
    \caption{Paediatric lupus application. Left: Classical type I error (true paediatric odds ratio = 1) and power (true paediatric odds ratio = 1.6) for the Bayesian study designs using robust mixture analysis prior and or vague analysis prior. Right: Enlargement of PoS curves for both analysis priors focusing on null or harmful values of the paediatric odds ratio, with adult and robust null design priors overlaid.}
    \label{fig:Benlysta_OC}
\end{figure}
\end{center}

Figure \ref{fig:Benlysta_OC} (left) and Table \ref{tab:Benlysta_OC} (top row) present the classical frequentist operating characteristics of the designs under the two different analysis priors. In contrast to Case Study 1 (where borrowing was on the placebo response and the pointwise classical type I error depended on the drift between the true placebo response and the historical placebo data), there is only one value of the classical type I error for a given BDB design when borrowing prior information directly on the treatment contrast. Therefore, the type I error and power do not vary with the true paediatric odds ratio, on the x-axis of Figure \ref{fig:Benlysta_OC}, but they correspond to selected values on this axis.

The prior information from adult data supports a positive treatment effect. Since, by definition, the type I error is evaluated by assuming the true treatment effect is null (i.e. a true odds ratio equal to 1), then it is calculated under a scenario where the prior is in conflict with the null treatment effect, resulting in a large inflation of the type I error with the BDB design using the robust adult prior ($33\%$, as compared to $2.5\%$ for the design with vague analysis prior that does not borrow adult information). As indicated in Section \ref{sec:considerations_treat}, a strict control of the type I error while borrowing on historical data is not possible in this context. Figure \ref{fig:Benlysta_OC} (left) also illustrates the `flip side' of borrowing the adult prior information in terms of the power gain (blue lines): for example, the probability of meeting the trial success criteria if the true paediatric odds ratio is 1.6 is 77\% for the BDB design and only 21\% for the Bayesian design with vague analysis prior. 
\subsection{Average type I error}

For the Bayesian design using each of the two analysis priors, Table \ref{tab:Benlysta_OC} (middle row) shows the average type I error (\ref{eq:typeIdelta}) evaluated under two alternative design priors:
\begin{enumerate}
\item Truncated adult design prior, chosen to be the normalised truncated lower tail ($\leq 0$) of the posterior from the pooled adult studies under an initial improper prior. 
\item Truncated robust design prior, chosen to be the normalised truncated lower tail ($\leq 0$) of the robust adult analysis prior.  
\end{enumerate}
Figure \ref{fig:Benlysta_OC} (right) shows these two design priors superimposed on the PoS curves for each of the analysis priors. Recall that the average type I error is calculated by integrating each of the PoS curves {\it wrt} the relevant design prior. From Figure \ref{fig:Benlysta_OC} (right), we see that the adult null design prior is highly concentrated near the null odds ratio of 1, with negligible weight on values for the odds ratio below 0.9. As a consequence, the average type I error under this design prior is very close to the frequentist type I error values (which can be thought of as the `average' PoS under a point mass design prior on odds ratio = 1). By contrast, the robust null design prior has a very heavy left tail, and is approximately uniform over log odds ratios in the range $(-\infty, log(1))$. This results in much lower average type I errors for both analysis models compared to under the null adult design prior. Such a heavy tail may be viewed as unreasonable for the null design prior, and so other options could be considered, such as truncating the lower tail of the null design prior at a plausible value \cite{psioda2019}. However judgements about what this value should be may be hard to elicit.

\subsection{Other metrics}
In the context of a paediatric bridging study, where there is scientific reason to expect the treatment effect in children to be similar to that demonstrated in adults, the prior probability of the null effect being true is expected to be low. Indeed, if there were a high prior probability of such a large drift between the true paediatric effect and the adult evidence, then a bridging strategy would not seem to be an option in the first place. 

Using the Bayesian framework, the prior probability of treatment benefit in children can be formally quantified, and is one of the metrics identified by Pennello and Thompson (2007) \cite{pennello2007} and Travis et al (2023) \cite{travis:2023} as being helpful for evaluating Bayesian designs in regulatory settings. For the paediatric lupus example, the prior probability of efficacy under the vague analysis prior is 50\%, which represents a position of clinical equipoise, but could be considered overly pessimistic in a setting where there is already confirmatory evidence of efficacy in adults. Under the robust mixture prior, this probability is 85\%, which reflects the available adult evidence and known similarities and differences between adults and children, whilst still leaving open the non-negligible possibility (15\%) that the drug is not effective in children. By contrast, if the adult prior were to be used directly as the analysis prior without any down-weighting, the prior probability of efficacy in children would be $>$99.9\%. This prior probability is greater than the decision threshold (which requires at least 97.5\% probability of efficacy) and so, if this analysis prior were considered justified, then an efficacy trial in children may not be considered necessary.              

The above discussion about prior probability of efficacy focuses on the prior information being used in the \textit{analysis} of the paediatric trial. It is also helpful to consider the probability of efficacy (or conversely, the probability of a null or harmful treatment effect) from the perspective of the {\it design} prior. As proposed in Section \ref{sec:considerations_treat}, in settings where there is reliable and relevant evidence favouring a non-null treatment effect, a useful metric is the (pre-posterior) probability of actually drawing a false positive conclusion (\ref{eq:uncondFP}), which we showed is equivalent to the average type I error multiplied by the probability of the null being true under a chosen design prior. The bottom part of Table \ref{tab:Benlysta_OC} reports these two probabilities, along with the upper bound on the probability of declaring a false positive result (given by the frequentist type I error multiplied by the probability of the null being true, as in (\ref{eq:upperbound})) under two different design priors representing the untruncated versions of the adult and robust design priors that were used to calculate the average type I error.  
\begin{table}[h]
\small\sf\centering
    \begin{center}
    \begin{tabular}{llrr}
        \toprule
  \textbf{Metric} & \multicolumn{1}{c}{\textbf{Analysis prior for}} & \multicolumn{2}{c}{\textbf{Design prior for}} \\
  & \multicolumn{1}{c}{\textbf{treatment difference}} & \multicolumn{2}{c}{\textbf{treatment difference}} \\ \hline \hline
 & & \multicolumn{2}{c}{\textit{No design prior}} \\ \cline{3-4}
 \textit{Frequentist type I error} & Vague & \multicolumn{2}{c}{2.5\%} \\
 \textit{(1-sided)} & Robust mixture & \multicolumn{2}{c}{33.2\%} \\ \hline 
 & & \textit{Truncated} & \textit{Truncated} \\ 
  & & \textit{adult} & \textit{robust mixture} \\ \cline{3-4}
 \textit{Average type I error} & Vague & 2.1\% & 0.1\%\\
                       &  Robust mixture & 30.8\% & 2.5\%\\ \hline 
 & & \textit{Adult} & \textit{Robust mixture} \\ \cline{3-4}
 \textit{Prior probability of}  & --- & 0.004\% & 15.003\%\\
 \textit{no treatment benefit}  & & & \vspace{2mm} \\ 
 \textit{Probability of actually} & Vague & $<$0.001\% & 0.015\%\\
 \textit{declaring false +ve result} &  Robust mixture & 0.001\% & 0.375\%\vspace{2mm} \\ 
 \textit{Upper bound on probability of} & Vague & $<$0.001\% & 0.375\%\\
 \textit{declaring false +ve result} &  Robust mixture & 0.001\% & 4.982\%  \\ \hline 
    \end{tabular}
    \caption{Paediatric lupus application. Operating characteristics of the Bayesian study designs using different analysis and design priors} 
    \label{tab:Benlysta_OC}
   \end{center}
\end{table}
 
The pre-posterior probability of actually declaring a false positive result, and the upper bound on this probability, are negligible for both analysis models under the adult design prior, reflecting the fact that the probability of no treatment benefit (i.e.~the null being true) under this prior is only 0.004\%. Even under the robust design prior, which has 15\% probability of the null being true, the chances of actually declaring a false positive result are less than 1\% for both analysis priors. The upper bound on this probability is somewhat higher, but still below 5\%, which may be considered a reasonable level of risk in challenging settings such a paediatrics.

\section{Discussion}
Bayesian clinical trials that leverage historical, or more generally study-external, data have become increasingly popular over time, offering a new toolbox in drug development. Many fully Bayesian clinical studies that leveraged historical data have been conducted already, and some of them were published in top-tier clinical journals \cite{Baeten2013,Boehm2020, Richeldi2022}. However, the vast majority of these studies were either done in early development (phase I/II), post-approval (phase IV) or undertaken by academic groups. The few examples used for registration purposes were essentially for paediatric indications or in rare and ultra-rare diseases \cite{Goring2019}. One likely reason for this is that strict control of the classical (frequentist) type I error, which is not possible when leveraging historical data, has proven as a sticking point in discussions with regulators. However, as the frequentist (long-run) interpretation of probability is not the only way to interpret probabilities, likewise in this paper we argue that the frequentist viewpoint on type I error is not the only one we should adopt. Instead, we suggest that Bayesian metrics should be used for Bayesian designs. While the concept of assurance \cite{OHagan2005} -- or Bayesian average power \cite{chuang-stein2017} -- has become instrumental nowadays to support decision-making, we argue that the average type I error, which is its equivalent under the null hypothesis, is also a relevant metric to inform decision-makers about the risk of a false positive result associated with a Bayesian design. In designs where information is borrowed on the treatment contrast, we further recommend calculation of the probability of actually declaring a false positive result, to provide further insights on the design performance. Altogether, these metrics provide a comprehensive and reliable way to assess the properties of Bayesian studies as illustrated in the two case studies that we presented.

An important feature of our work is that it relies on a willingness to adopt the Bayesian approach also for assessing the risk of, e.g., declaring a treatment as effective which in reality is ineffective. Consequently, it is a necessity to agree on \textit{analysis} and \textit{design} priors, with the latter being used to assess the operating characteristics of the trial design under various scenarios (`what if' situations). This is a somewhat elaborate task, in particular since many professionals may still be unfamiliar with it, given that assessing the classical type I error rate has no comparable requirement (the null hypothesis is, in a sense, trivial). Thus, more upfront discussion and alignment may be needed before an agreement is reached. Furthermore, when borrowing on the treatment contrast, the choice of the {\it null} design prior to evaluate the unconditional type I error can be particularly challenging. By virtue of its nature, borrowing on the treatment contrast usually implies that the treatment has some effect, i.e., there is possibly substantial evidence against the null hypothesis. Therefore, we embraced ideas outlined in the work by Spiegelhalter and Friedman \cite{spiegelhalter1986} and Chuang-Stein and Kirby \cite{chuang-stein2017} to calculate certain joint (or pre-posterior) probabilities that reflect the probability of the null hypothesis actually being true. Whilst these joint probabilities also require specification of a design prior, this prior is not restricted to be consistent with the null hypothesis. To further simplify specification of the design prior in this setting, we also proposed using the upper bound on the pre-posterior probability of actually drawing a false positive conclusion. This just requires specification of the tail area probability of the null being true under the design prior (i.e.~the shape of the null tail is irrelevant and so it can be thought of as a point mass at the null), which is then multiplied by the classical type I error.

In their discussion of regulatory perspectives on informative Bayesian methods for paediatric efficacy trials, Travis et al \cite{travis:2023} comment on the importance of taking into consideration the likelihood of the true parameter value being in a particular region when assessing the type I error. They also note that an important factor for regulators is the ability to implement consistent standards across studies. We argue that use of Bayesian metrics involving explicit definition of a design prior can help to address both these requirements. The design prior provides a mechanism to pre-specify and make explicit what assumptions are being made about the chances of the true parameter values of interest (e.g.~true control response; true treatment contrast) being in any particular region when evaluating the operating characteristics of a clinical trial design. The design prior would need to be agreed between sponsors and regulators on a case by case basis to reflect plausible assumptions about the disease and the treatment effects. However, conditional on an agreed design prior, regulators could then require control of appropriate Bayesian metrics, such as those we discuss here, at a consistent level across products. For example, designs that borrow prior information on the control arm could be required to control average type I error at a prescribed level that is consistent for all products in a certain class or disease area or category of unmet need. For designs that borrow information on the treatment contrast, one approach could be to require the upper bound on the pre-posterior probability of actually making a type I error to be consistent between different products. As already noted, agreeing a design prior in this case reduces to requiring sponsors and regulators to agree on the prior probability of the true treatment effect being null. The sponsor could then optimise the trial design balancing sample size against the amount of external information to borrow in order to maintain the upper bound on a false positive outcome at the agreed level. Such an approach could also address Travis et al \cite{travis:2023}'s concern that using the same classical type I error rate across products will force less borrowing for more effective products. Whilst a more effective product would, indeed, have a higher conditional type I error if borrowing the same amount of prior information as a less effective product, we would expect the design prior for the more effective product to have a lower probability of the null being true. Hence the pre-posterior probability of actually making a type I error would not necessarily be higher for the more effective product. 

It is worth noting that more than one design prior may be specified if desired, in order to reflect a range of stakeholder opinions. Thus we could envisage a situation where regulators agree two different design priors, representing, say optimistic and sceptical judgements about the likelihood of the true parameters being in certain regions. Consistent thresholds for relevant Bayesian metrics could be set for each type of design prior, with a more stringent level of control being required under the optimistic design prior.    

One reason for differences between the historical prior and new trial data that could lead to an increased risk of erroneous conclusions in a Bayesian borrowing design is systematic variation in baseline prognostic factors. If these factors are measured in the historical and new data, this source of drift can be addressed through statistical modelling such as regression or propensity score weighting \cite{banbeta:2022, fu:2023}.  All the metrics presented in this paper extend naturally to this situation. The target parameter for information borrowing in the new study (i.e.~$\theta_c$ or $\delta$) is then the marginal \textit{covariate-adjusted} control treatment effect or treatment contrast, respectively. Since at the design stage, the exact distribution of covariates in the new study is usually unknown, an explicit assumption must be made about the expected covariate distribution. The metrics in Section 2 then require specification of a design prior to describe assumptions about the marginal covariate-adjusted treatment effect or treatment contrast in the new study for the specified covariate distribution.

The evaluation of designs that leverage historical data may look quite involved. However, it is noteworthy to mention that the complexity of many designs -- whether they leverage historical data or not -- often require simulations to `stress-test' assumptions and understand their operating characteristics. Therefore, it is probably fair to assume that the technical skills required to perform the evaluations of Bayesian designs as proposed in this work should not present a barrier anymore. The more challenging task might be to familiarize stakeholders with the concept of metrics that go beyond the (well-known) classical type I error \cite{EMA_ITF} . Explaining which risks the metrics help to assess and how they complement other existing metrics may thus be an important part when presenting the evaluation of a design that uses historical data. Concurrently, it is often very useful to discuss with key stakeholders so called \textit{data scenarios}, i.e., hypothetical data and accompanying results as they could actually occur in the study. This often helps people to get a very concrete impression of what could happen, as opposed to average statements as obtained from metrics. 

Finally, it might also be useful to remind ourselves that the fundamental questions around studies that leverage historical data have not changed over time. In his seminal work \textit{Designing for nonparametric Bayesian survival analysis using historical controls} \cite{vanryzin1980}, published in 1980, John van Ryzin nicely summarized that it is an elementary bias-variance trade-off question which ultimately should guide if -- and to what extent -- historical data are to be used. The strong focus on classical (frequentist) type I error control for pivotal studies, however, has shifted much of the considerations to the bias question only. It is thus important to re-consider what metrics should be used for evaluating designs that -- \textbf{by construction} -- aim at optimizing the bias-variance trade-off, such as Bayesian designs using historical data, or adaptive designs with treatment or population selection at interim \cite{Robertson2023}. Similarly, it is paramount to better understand how existing metrics could be used in a more principled way \cite{grieve2016,walley2021}. The current situation is in some sense similar to what happened when Stein \cite{stein1956} and James and Stein \cite{jamesstein1961} showed that the ``usual" least-squares estimator for the mean is admissible in dimensions three or higher. While the James-Stein estimator is biased, it outperforms the least-squares estimator with regards to mean squared error. Thus, a more holistic viewpoint was required from a metrics perspective to judge which estimator to prefer. Similarly, we conclude that a more holistic viewpoint is required to judge which study design serves the purpose of a registration trial best.

\section*{Conflict of interests}
Nicky Best is an employee of GSK and holds shares of GSK. Maxine Ajimi is an employee of AstraZeneca and holds shares of AstraZeneca. Beat Neuenschwander is an employee of Novartis and holds shares of Novartis. Gaëlle Saint-Hilary is president and single associate of Saryga. Simon Wandel is an employee of Novartis and holds shares of Novartis. 

\bibliographystyle{Bibstyle}
\bibliography{Bibliography}

\begin{thebibliography}{10}
\providecommand{\url}[1]{\texttt{#1}}
\providecommand{\urlprefix}{URL }
\expandafter\ifx\csname urlstyle\endcsname\relax
  \providecommand{\doi}[1]{DOI:\discretionary{}{}{}#1}\else
  \providecommand{\doi}{DOI:\discretionary{}{}{}\begingroup
  \urlstyle{rm}\Url}\fi
\providecommand{\eprint}[2][]{\url{#2}}

\bibitem{Petrou2012}
Petrou S.
\newblock Rationale and methodology for trial-based economic evaluation.
\newblock \emph{Clin Invest} 2012; 2(12): 1191--1200.
\newblock Doi: 10.4155/cli.12.121.

\bibitem{Moore2018}
Moore T, Zhang H, Anderson G et~al.
\newblock {Estimated Costs of Pivotal Trials for Novel Therapeutic Agents
  Approved by the US Food and Drug Administration, 2015--2016}.
\newblock \emph{Jama Intern Med} 2018; 178(11): 1451--1457.
\newblock Doi:10.1001/jamainternmed.2018.3931.

\bibitem{FDAClinicalEvidence1998}
{Food and Drug Administration}.
\newblock {Providing Clinical Evidence of Effectiveness for Human Drugs and
  Biological Products (Guidance for Industry)} 1998; Retrieved from
  https://www.fda.gov/media/71655/download.

\bibitem{FDADemSubEvidence2019}
{Food and Drug Administration}.
\newblock {Demonstrating Substantial Evidence of Effectiveness for Human Drug
  and Biological Products (Guidance for Industry)} 2019; Retrieved from
  https://www.fda.gov/media/133660/download.

\bibitem{FDAAdaptDesign2019}
{Food and Drug Administration}.
\newblock {Adaptive Designs for Clinical Trials of Drugs and Biologics
  (Guidance for Industry)} 2019; Retrieved from
  https://www.fda.gov/media/78945/download.

\bibitem{Goring2019}
Goring S, Taylor A, M\"uller K et~al.
\newblock {Characteristics of non-randomised studies using comparisons with
  external controls submitted for regulatory approval in the US and Europe: a
  systematic review}.
\newblock \emph{BMJ open} 2019; 9(2): e024895.
\newblock Doi:10.1136/bmjopen-2018-024895.

\bibitem{schmidli2014}
Schmidli H, Gsteiger S, Roychoudhury S et~al.
\newblock Robust meta-analytic-predictive priors in clinical trials with
  historical control information.
\newblock \emph{Biometrics} 2014; 70(4): 1023--1032.
\newblock Doi: 10.1111/biom.12242.

\bibitem{FDAInteractingComplex2019}
{Food and Drug Administration}.
\newblock {Interacting with the FDA on Complex Innovative Trial Designs for
  Drugs and Biological Products (Guidance for Industry)} 2020; Retrieved from
  https://www.fda.gov/media/130897/download.

\bibitem{viele2014}
Viele K, Berry S, Neuenschwander B et~al.
\newblock Use of historical control data for assessing treatment effects in
  clinical trials.
\newblock \emph{Pharm Stat} 2014; 13(1): 41--54.
\newblock Doi: 10.1002/pst.1589.

\bibitem{koppschneider2020}
Kopp-Schneider A, Calderazzo S and Wiesenfarth M.
\newblock Power gains by using external information in clinical trials are
  typically not possible when requiring strict type i error control.
\newblock \emph{Biometrical Journal} 2020; 62(2): 361--374.
\newblock Doi: 10.1002/bimj.201800395.

\bibitem{psioda2019}
Psioda M and Ibrahim J.
\newblock Bayesian clinical trial design using historical data that inform the
  treatment effect.
\newblock \emph{Biostatistics} 2019; 20(3): 400--415.
\newblock Doi: 10.1093/biostatistics/kxy009.

\bibitem{pennello2007}
Pennello G and Thompson L.
\newblock Experience with reviewing bayesian medical device trials.
\newblock \emph{Journal of Biopharmaceutical Statistics} 2007; 18(1): 81--115.
\newblock Doi: 10.1080/10543400701668274.

\bibitem{walley2015}
Walley RJ, Smith CL, Gale JD et~al.
\newblock Advantages of a wholly bayesian approach to assessing efficacy in
  early drug development: a case study.
\newblock \emph{Pharm Stat} 2015; 14(3): 205--15.
\newblock Doi: 10.1002/pst.1675.

\bibitem{roychoudhury2018}
Roychoudhury B, Scheuer N and Neuenschwander B.
\newblock {Beyond p-values: A phase II dual-criterion design with statistical
  significance and clinical relevance}.
\newblock \emph{Clinical Trials} 2018; 15(5): 452--461.
\newblock Doi: 10.1177/1740774518770661.

\bibitem{OHagan2005}
O'Hagan A, Stevens JW and Campbell MJ.
\newblock Assurance in clinical trial design.
\newblock \emph{Pharm Stat} 2005; 4: 187--201.
\newblock Doi: 10.1002/pst.175.

\bibitem{grieve2016}
Grieve A.
\newblock Idle thoughts of a `well-calibrated' {B}ayesian in clinical drug
  development.
\newblock \emph{Pharmaceutical Statistics} 2016; 15: 96--108.
\newblock Doi: 10.1002/pst.1736.

\bibitem{viele:2018}
Viele K, Mundy L, Noble R et~al.
\newblock Phase 3 adaptive trial design options in treatment of complicated
  urinary tract infection.
\newblock \emph{Pharmaceutical Statistics} 2018; 17: 811--22.
\newblock Doi: 10.1002/pst.1892.

\bibitem{lim:2019}
Lim J, Wang L, Best N et~al.
\newblock Minimizing patient burden through the use of historical subject-level
  data in innovative confirmatory clinical trials: Review of methods and
  opportunities.
\newblock \emph{Therapeutic Innovation \& Regulatory Science} 2019; 52:
  546--59.
\newblock Doi: 10.1177/2168479018778282.

\bibitem{spiegelhalter2004}
Spiegelhalter D, Abrams K and Myles J.
\newblock \emph{Bayesian approaches to clinical trials and health-care
  evaluation}.
\newblock Chichester, UK: John Wiley \& Sons Ltd, 2004.

\bibitem{wang2002}
Wang F and Gelfand A.
\newblock A simulation-based approach to bayesian sample size determination for
  performance under a given model and for separating models.
\newblock \emph{Statistical Science} 2002; 17(2): 193--208.

\bibitem{psioda2020}
Psioda M and Xue X.
\newblock A bayesian adaptive two-stage design for pediatric clinical trials.
\newblock \emph{Journal of Biopharmaceutical Statistics} 2020; 30(6):
  1091--1108.
\newblock Doi: 10.1080/10543406.2020.1821704.

\bibitem{spiegelhalter1986}
Spiegelhalter D and Friedman L.
\newblock A predictive approach to selecting the size of a clinical trial,
  based on subjective clinical opinion.
\newblock \emph{Statistics in Medicine} 1986; 5(1): 1--13.
\newblock Doi: 10.1002/sim.4780050103.

\bibitem{chuang-stein2017}
Chuang-Stein C and Kirby S.
\newblock \emph{Quantitative Decisions in Drug Development}.
\newblock Cham, Switzerland: Springer, 2017.

\bibitem{Hueber2012}
Hueber W, Sands BE, Lewitzky S et~al.
\newblock {Secukinumab, a human anti-IL-17A monoclonal antibody, for moderate
  to severe Crohn's disease: unexpected results of a randomised, double-blind
  placebo-controlled trial}.
\newblock \emph{Gut} 2012; 61(12): 1693--1700.
\newblock Doi:10.1136/gutjnl-2011-301668.

\bibitem{RBesT}
Weber S, Neuenschwander B, Schmidli H et~al.
\newblock \emph{RBesT: R Bayesian Evidence Synthesis Tools}, 2020.
\newblock
  \urlprefix\url{https://cran.r-project.org/web/packages/RBesT/index.html}.
\newblock R package version 1.6-1.

\bibitem{best1976}
Best W, Becktel J, Singleton J et~al.
\newblock Development of a crohn's disease activity index. national cooperative
  crohn's disease study.
\newblock \emph{Gastroenterology} 1976; 70(3): 439--444.

\bibitem{Neuenschwander2010}
Neuenschwander B, Capkun-Niggli G, Branson M et~al.
\newblock Summarizing historical information on controls in clinical trials.
\newblock \emph{Clinical Trials} 2010; 7(1): 5--18.
\newblock Doi: 10.1177/1740774509356002.

\bibitem{FDABLA}
{Food and Drug Administration}.
\newblock {BLA 125370/s-064 and BLA 761043/s-007 Multi-disciplinary Review and
  Evaluation} 2018; Retrieved from https://www.fda.gov/media/127912/download.

\bibitem{travis:2023}
Travis J, Rothman M and Thomson A.
\newblock Perspectives on informative {B}ayesian methods in pediatrics.
\newblock \emph{Journal of Biopharmaceutical Statistics} 2023; 15: 96--108.
\newblock Doi: 10.1080/10543406.2023.2170405.

\bibitem{Baeten2013}
Baeten D, Baraliakos X, Braun J et~al.
\newblock Anti-interleukin-17a monoclonal antibody secukinumab in treatment of
  ankylosing spondylitis: a randomised, double-blind, placebo-controlled trial.
\newblock \emph{Lancet} 2013; 382(9906): 1705--1713.
\newblock Doi: 10.1016/S0140-6736(13)61134-4.

\bibitem{Boehm2020}
B\"ohm M, Kario K, Kandzari D et~al.
\newblock Efficacy of catheter-based renal denervation in the absence of
  antihypertensive medications (spyral htn-off med pivotal): a multicentre,
  randomised, sham-controlled trial.
\newblock \emph{Lancet} 2020; 395(10234): 1444--1451.
\newblock Doi: 10.1016/S0140-6736(20)30554-7.

\bibitem{Richeldi2022}
Richeldi L, Azuma A, Cottin V et~al.
\newblock Trial of a preferential phosphodiesterase 4b inhibitor for idiopathic
  pulmonary fibrosis.
\newblock \emph{N Engl J Med} 2022; 386(23): 2178--2187.
\newblock Doi: 10.1056/NEJMoa2201737.

\bibitem{banbeta:2022}
Banbeta A, Lesaffre E and van Rosmalen J.
\newblock The power prior with multiple historical controls for the linear
  regression model.
\newblock \emph{Pharm Stat} 2022; 21: 418--38.
\newblock Doi: 10.1002/pst.2178.

\bibitem{fu:2023}
Fu C, Pang H, Zhou S et~al.
\newblock Covariate handling approaches in combination with dynamic borrowing
  for hybrid control studies.
\newblock \emph{Pharm Stat} 2023; Doi: 10.1002/pst.2297.

\bibitem{EMA_ITF}
{European Special Interest Group on Historical Data}.
\newblock {A framework for evaluation of Bayesian Dynamic Borrowing designs in
  pivotal studies}.
\newblock \emph{{PSI conference}} 2022; Retrieved from
  https://www.psiweb.org/sigs-special-interest-groups/historical-data.

\bibitem{vanryzin1980}
Van~Ryzin J.
\newblock Designing for nonparametric bayesian survival analysis using
  historical controls.
\newblock \emph{Cancer Treat Rep} 1980; 64(2-3): 503--506.

\bibitem{Robertson2023}
Robertson D, Choodari-Oskooei B, Dimairo M et~al.
\newblock {Point estimation for adaptive trial designs I: A methodological
  review}.
\newblock \emph{Statistics in Medicine} 2023; 42(2): 122--145.
\newblock Doi: 10.1002/sim.9605.

\bibitem{walley2021}
Walley R and Grieve A.
\newblock {Optimising the trade-off between type I and II error rates in the
  Bayesian context}.
\newblock \emph{Pharm Stat} 2021; 20(4): 710--720.
\newblock Doi: 10.1002/pst.2102.

\bibitem{stein1956}
Stein C.
\newblock Inadmissibility of the usual estimator for the mean of a multivariate
  distribution.
\newblock \emph{Proc Third Berkeley Symp Math Statist Prob} 1956; 1: 197--206.

\bibitem{jamesstein1961}
James W and Stein C.
\newblock Estimation with quadratic loss.
\newblock \emph{Proc Fourth Berkeley Symp Math Statist Prob} 1961; 1: 361--379.

\end{thebibliography}

\end{document}